\documentclass[aps,prl,twocolumn,superscriptaddress]{revtex4-1}
\usepackage{amsmath}
\usepackage{bm}
\usepackage{graphicx}
\usepackage{cancel}
\usepackage{subfigure}
\usepackage{mathrsfs}
\usepackage{float}
\usepackage{amssymb}

\begin{document}

\title{Generation of a squeezed state of an oscillator by stroboscopic back-action-evading measurement}

\date{\today}

\author{G. Vasilakis}
\thanks{G. Vasilakis and H. Shen contributed equally to this work.}
\affiliation{Niels Bohr Institute, Copenhagen University,\\
Blegdamsvej 17, 2100 Copenhagen, Denmark}

\author{H. Shen}
\thanks{G. Vasilakis and H. Shen contributed equally to this work.}
\affiliation{Niels Bohr Institute, Copenhagen University,\\
Blegdamsvej 17, 2100 Copenhagen, Denmark}

\author{K. Jensen}
\affiliation{Niels Bohr Institute, Copenhagen University,\\
Blegdamsvej 17, 2100 Copenhagen, Denmark}

\author{M. Balabas}
\affiliation{Department of Physics, St. Petersburg State University,\\
Universitetskii pr. 28, 198504 Staryi Peterhof, Russia}
\affiliation{Niels Bohr Institute, Copenhagen University,\\
Blegdamsvej 17, 2100 Copenhagen, Denmark}

\author{D. Salart}
\affiliation{Niels Bohr Institute, Copenhagen University,\\
Blegdamsvej 17, 2100 Copenhagen, Denmark}

\author{B. Chen}
\affiliation{Niels Bohr Institute, Copenhagen University,\\
Blegdamsvej 17, 2100 Copenhagen, Denmark}
\affiliation{Quantum Institute of Atom and Light, State Key Laboratory of Precision Spectroscopy, Department of Physics, East China Normal University,\\
Shanghai 200062, P. R. China.}

\author{E. S. Polzik}
\email{polzik@nbi.ku.dk}
\affiliation{Niels Bohr Institute, Copenhagen University,\\
Blegdamsvej 17, 2100 Copenhagen, Denmark}

\begin{abstract}
Continuous observation on an oscillator is known to result in quantum back-action which limits the knowledge acquired by the measurement.
A careful balance between the information obtained and the back-action disturbance leads to a limit known as the standard quantum limit.
The means to surpass this limit by modulating the measurement strength with the period proportional to half period of the oscillation has been proposed decades ago \cite{BraginskyPaper, CavesThornePRL,QNDBraginskyScience}.  Such modulated or stroboscopic observation leading to a squeezed state of one quadrature of the oscillator motion with the quantum noise below that of the zero-point fluctuations has been a long-standing goal. Here, we report on the generation of a quadrature-squeezed state of an oscillator by stroboscopic back-action evading measurement. The oscillator is the collective spin of an atomic ensemble precessing in magnetic field. It is initially prepared in nearly the ground state with an average thermal occupancy number $0.08 \pm 0.01$.  The oscillator is coupled to the optical mode of a cavity, and the cavity output field detected with polarization homodyning serves as the meter.  A back-action-evading measurement is performed by stroboscopically modulating the intensity of the light field at twice the Larmor frequency,
resulting in a squeezed state conditioned on the light-polarization measurement with $2.2 \pm 0.3$ dB noise reduction below the zero-point fluctuations for the measured quadrature. The demonstrated squeezing holds promise for metrological advantage in quantum sensing.
\end{abstract}

\maketitle

The Heisenberg uncertainty sets the limit of how precisely two non-commuting variables, such as the canonical position and momentum with the commutation relation $[\hat{X_0},\hat{P_0}]=i$, can be specified simultaneously; however, there are no physical limitations on determining the value of an individual variable.
For an oscillator with frequency $\Omega$ the position variable in the laboratory frame is $\hat{X}=\hat{X_0} \cos(\Omega t)+\hat{P_0} \sin(\Omega t)$ where $\hat{X_0},\hat{P_0}$ are the rotating frame variables. Quantum states for which a canonical variable  has reduced uncertainty with respect to the oscillator ground state, for example, $\text{Var} ( \hat{X_0} ) <1/2$, are called squeezed states (SS).
Squeezed states for matter oscillators have been first demonstrated for motion of a single ion \cite{WinelandSqueezingIons} and later for magnetic oscillators by electron-nucleous entanglement \cite{FPPRLNuclElectSq,PolzikRMP} and by spin-spin interaction \cite{GrossNature, TreutleinNatureEntangl}.

An intriguing approach towards generation of an SS in an oscillator is a stroboscopic quantum non demolition (QND) or back-action evading measurement proposed in \cite{BraginskyPaper, CavesThornePRL,QNDBraginskyScience}.  If a meter is coupled to the oscillator with an interaction Hamiltonian $\hat{H}\propto \hat{X}$ and if the measurement strength is maximized at times $t=0, \pi/\Omega,... n\pi/\Omega$ the noise of the meter does not couple to $\hat{X_0}$ and the readout of this QND measurement yields a SS of the oscillator with a degree of squeezing defined as $\xi^2=2 \text{Var}( \hat{X_0}) <1$.
Another version of this approach based on harmonic modulation of the measurement strength which works similarly to the stroboscopic measurement  has been proposed in \cite{CMJNewJPhys}. Previous attempts to implement this approach with a magnetic spin oscillator \cite{RomalisStroboscopicPRL} and a mechanical oscillator \cite{SchwabScience} demonstrated reduction in the back-action noise but did not result in an SS due to the insufficient strength of the QND measurement compared to the decoherence caused by the environment. In a separate line of work back-action evasion has been demonstrated for a joint state of two oscillators \cite{HamArchive,PolzikRFPRL}. There a continuous QND measurement on both oscillators one of which has an effective negative mass has been shown to generate an entangled state of the two oscillators.

A key requirement for achieving the SS is that the decoherence due to interaction with the environment is reduced to the extent that the oscillator maintains its quantum state for the time longer than a QND measurement takes.  Every decoherence event is linked with loss of information about the measured variable and with the decay to the ground or thermally excited state of the oscillator depending on the decoherence mechanism.  In this work this requirement is met by firstly enhancing the rate of the optical QND interaction by placing the spin ensemble in an optical cavity, and secondly by placing the spins in a designed spin-protecting environment.

\begin{figure}
  \includegraphics[scale=0.5]{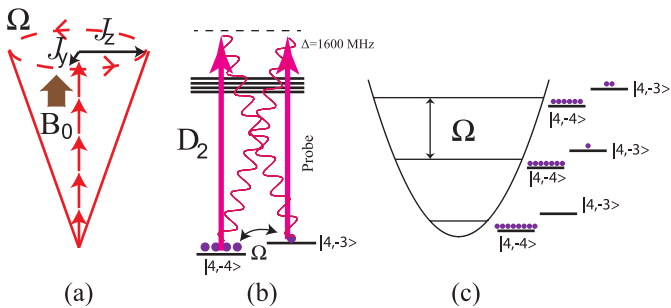}
  \caption{Magnetic oscillator. (a) The collective spin of an ensemble of atoms with macroscopic orientation along the  magnetic field $B_0$ precesses around  $B_0$ with a Larmor frequency $\Omega$. Normalized quantized values of the projections $\hat{J}_{y0,z0}$ in the rotating frame are the canonical variables for this oscillator.  (b) The energy diagram of an atomic constituent of the oscillator. In the ground state of the oscillator atoms are in the lower energy state $|4,-4\rangle$. Raman scattering of photons (dashed lines) driven by the input light (solid arrows) creates collective excitations in level $|4,-3 \rangle$ and corresponding quantum coherences $|4,-3 \rangle \langle 4,-4|$ responsible for $\hat{J}_{y,z}$ and oscillating at $\Omega$. (c) The $n$-th excited state of the oscillator corresponds to $n$ collective spin flips generated as in (b). A squeezed state is a coherent superposition of such number states (comments in the text).}
 \label{fig:SpinOscillator}
\end{figure}

The oscillator used in this study is a collective spin of an atomic ensemble precessing in a bias magnetic field (Fig.~\ref{fig:SpinOscillator}a). The collective spin components in the lab frame oscillate as $\hat{J}_{z/y}=\pm\hat{J}_{z0/y0}\cos(\Omega t)/\sin(\Omega t)\mp \hat{J}_{y0/z0}\sin(\Omega t)/\cos(\Omega t)$. For a macroscopic spin orientation $J_x$ along the magnetic field, the energy spectrum of the spin ensemble can be mapped onto that of a harmonic oscillator  \cite{PolzikRMP} with a resonance frequency set by the Larmor frequency of the spin precession (Fig.~\ref{fig:SpinOscillator}b,c). Mathematically this is expressed through the Holstein-Primakoff transformation \cite{HolsteinPrimakoff}. Canonical position and momentum operators can be defined through the collective symmetric spin observables in the rotating frame $[\hat{J}_{y0},\hat{J}_{z0}]=i J_x=i N_{\text{at}}F$ ($\hbar=1$) as: $\hat{X_0}= \hat{J}_{z0}/\sqrt{ | \langle {J}_x \rangle | }$, and $\hat{P_0}= \hat{J}_{y0}/\sqrt{ | \langle {J}_x \rangle | }$ where it is assumed that the collective spin is well oriented, so that the population of the end state with the magnetic quantum number $m_F=-F$ is close to the total number of spins $N_{\text{at}}$  and hence the macroscopic quantity $J_x$ is treated as a number rather than an operator.
The ground state of the harmonic oscillator corresponds to all atoms being in the $F=4,m_F=-4$ state and the absence of the ensemble coherence $|4,-3\rangle \langle 4,-4|$ (Fig.~\ref{fig:SpinOscillator}b). The mean number of excitations above the ground state can be evaluated as $\bar{n}=\text{Var}( \hat{X} ) + \text{Var} ( \hat{P} )-1$. An excitation with the creation operator $\hat{a}^\dag=(\hat{X}-i \hat{P})/\sqrt{2}$ corresponds to a quantum of excitation, also called a polariton, distributed symmetrically among all the atoms of the oscillator. The ground state of the oscillator belongs to the class of Coherent Spin States (CSS) characterized by $\text{Var}(\hat{J}_{y0})=\text{Var}(\hat{J}_{z0})=J_x/2=N_{\text{at}}F/2$ \cite{PolzikRMP}.

The quantum state of the spin-oscillator can be  created and probed through interaction with a light field.  In the limit of large probe detuning with respect to the atomic excited state hyperfine level (Fig.~\ref{fig:SpinOscillator}b), the light-spin interaction can be approximated by the QND-type Hamiltonian: $H_{\text{int}}=2 (\kappa/\sqrt{N_{\text{ph}}} ) \hat{S}_z \hat{X}$, where $N_{\text{ph}}$ is the number of photons in the pulse of duration $\tau$ and
$\hat{S}_z$ is the probe light Stokes operator in the circular basis
normalized so that for a coherent pulse Var$\left(S_z \right)=N_{\text{ph}}/(4\tau)$. The coupling constant $\kappa$ characterizes QND interaction strength, which depends on the atom-light detuning, the excited-state linewidth and $\kappa \propto \sqrt{N_{\text{at}} N_{\text{ph}}}\propto \sqrt{d_0 \eta_{\tau}}$  \cite{PolzikRMP}, where $d_0$ is the resonant optical depth of the atomic ensemble and $\eta_{\tau}$ is proportional to the fractional number of decoherence events during the measurement time \cite{Supplemental}.

\begin{figure}
  \includegraphics[scale=1]{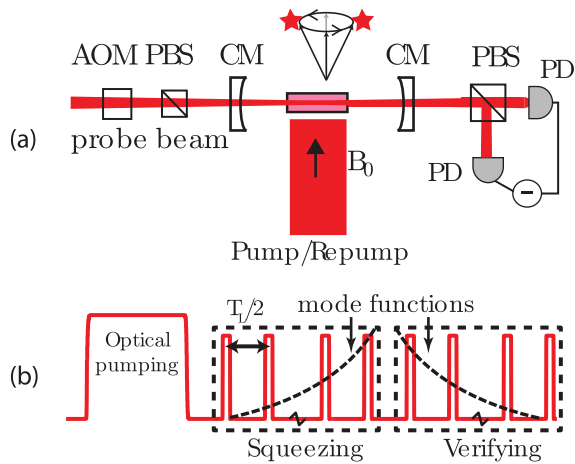}
  \caption{(a) Outline of the experimental setup. The acousto-optic modulator (AOM) generates light pulses for stroboscopic measurement of the atomic spin oscillator placed in an optical cavity formed by mirrors (CM). Quantum measurement of the polarization state of light is performed with polarization beamsplitter (PBS) and two photodetectors (PD).  The spin oscillator precesses in $B_0$. The projection of the spin along the probe direction is subject to stroboscopic QND measurements every half period of oscillation as indicated with stars. (b) Pulse sequence for generating and demonstrating conditional squeezing of an oscillator with stroboscopic QND measurement.}
   \label{fig:Experimental Setup}
\end{figure}

For stroboscopic probing the relevant observable for the oscillator is the harmonic quadrature that evolves in phase with the modulation (chosen to be the cosine quadrature): $\hat{x} = (1/T D)\int \! \mathrm{d}t \, \hat{X} (t) \phi(t) \cos \left( \Omega t \right)  $, where $T=2 \pi / \Omega$ is the oscillator period, $D$ is the duty cycle of the probe, $\phi(t)$ is a rectangular-pulse shaping function of unit amplitude, following the temporal evolution of the probe power, and the integration extends over one oscillator period. In the limit of zero duty cycle: $\hat{x}=\hat{X}_0$. The measurement record is the $\cos \Omega t$ Fourier component of the photocurrent integrated over the pulse length:  $\hat{S}_{y,\tau} = \int_0^\tau \! \mathrm{d}t \,   \hat{S}_y(t)  u(t) \cos \left( \Omega t \right) $, where $\hat{S}_y$ is the Stokes operator in the linear basis measured with polarization homodyning as shown in Fig.~\ref{fig:Experimental Setup} and $u$ is a mode function of unit energy over the pulse length that weights the measurement data according to the decoherence rate (see below).
The evolution of variances due to the Hamiltonian interaction is described by (for $u=1$)\cite{Supplemental}:
\begin{align}
\text{Var} (\hat{S}_{y,\tau}) & = \frac{\mathcal{B} N_{\text{ph}}}{8} \left[ 1+   \tilde{\kappa}^2 \text{Var} (\hat{x}_{\text{in}})  + \mathcal{C} \frac{\tilde{\kappa}^4}{3}  \right],  \label{eq:PolarimetryVariance} \\
\text{Var} (\hat{x}_{\text{out}}) & = \text{Var} (\hat{x}_{\text{in}}) +\mathcal{C} \tilde{\kappa}^2, \label{eq:xVariance}
\end{align}
where the subscripts (in), (out) indicate operators at the start and end of the interaction respectively, $\mathcal{B}=1+\text{sinc}(\pi D)$, $\tilde{\kappa}=\kappa \sqrt{\mathcal{B}}$ is the modified coupling constant, and  $\mathcal{C} \in \left[0,1 \right]$ quantifies the coupling of the probe noise to the observable. It can be shown \cite{Supplemental} that for a stroboscopic probe the back-action coupling constant is given by:
\begin{equation}
\mathcal{C}=\frac{1-\text{sinc}(\pi D)}{ 1+\text{sinc}(\pi D) }. \label{eq:BANcoupling}
\end{equation}
For a QND measurement $\mathcal{C}=0$, while for a continuous probing $\mathcal{C}=1$. In Eq. \ref{eq:PolarimetryVariance}, the first term on the right-hand side specifies the imprecision due to the quantum noise of the meter (shot noise of light), whereas the second and third terms describe the oscillator input noise and back-action noise, respectively. In the following, the measurement noise  associated with the oscillator in photon shot noise units will be denoted collectively with $\text{Var} (\hat{x}_\text{m})  =\tilde{\kappa}^2 \text{Var} (\hat{x}_{\text{in}})  + \mathcal{C} \frac{\tilde{\kappa}^4}{3}+\eta_{\tau}$, where $\eta_\tau$ expresses the uncertainty increase due to decoherence.
Conditionally on a QND ($\mathcal{C}=0$) realization of $\hat{S}_{y,\tau}$ the oscillator evolves to a quantum state with reduced position noise: $\text{Var}(\hat{x}_{\text{out}} | \hat{S}_{y,\tau} ) = \text{Var} (\hat{x}_{\text{out}}) - \text{Cov}^2(\hat{x}_{\text{out}},\hat{S}_{y,\tau})/ \text{Var} (\hat{S}_{y,\tau}) $, with Cov denoting the covariance. The quantum filtering then leads to conditional squeezing described by \cite{Supplemental}:
\begin{equation}
\xi^2 = \frac{\text{Var}(\hat{x}_{\text{out}} | \hat{S}_{y,\tau} )}{\langle \hat{x}^2 \rangle_0}=\frac{1}{1+\tilde{\kappa}^2}+ \eta_\tau
\end{equation}
where $\langle \hat{x}^2 \rangle_0$ is the imprecision in the ground state (zero point fluctuations). The decoherence term signifies the importance of the large optical depth $d_0$ in achieving high degrees of squeezing, as $\tilde{\kappa}^2 \propto d_0 \eta_{\tau}$ . We achieve an enhancement of the effective $d_0$ by enclosing the spin oscillator in an optical resonator (Fig.~\ref{fig:Experimental Setup}). Optimal squeezing can be achieved with an impedance matched cavity where the output coupler transmission $T_{\text{out}} \approx \mathcal{L} $, where $\mathcal{L}$ is the round-trip intensity loss \cite{Supplemental}.

The spin oscillator is realized in room-temperature, optically-pumped Cesium atoms, contained in a glass cell microchannel, $300 \mu \text{m} \times 300 \mu \text{m}$ in cross section and 1~cm in length. An alkene coating \cite{BalabasPRLCoating} deposited at the inner cell walls dramatically suppresses spin-relaxation due to the wall collisions. Atoms bounce off the walls and cross the optical mode cross section with the waist of $55~\mu$m approximately $5\times 10^3$ times before their quantum spin state decoheres in $10$~msec due to wall collisions. The atom-atom collision rate at the low Cs pressure used here is negligible. As the typical light pulse duration of $\approx 2$~msec is much greater than the atom transient time of  $\approx 1.5~ \mu$sec and the oscillator period (typical $\Omega\sim 380$~kHz)  the thermally moving atoms cross the optical mode many times in the same state and hence the detected optical mode couples to the symmetric spin mode (equivalently to the oscillator position $\hat{X}$). We emphasize that the thermal motion of the atoms does not affect the oscillator temperature, which is determined by the spin distribution. The microcell is placed inside a standing wave optical cavity with a finesse $\mathscr{F} \approx 17$, the single-pass losses in the cell windows of $6.5\%$ and the output coupler transmission of $80\%$, which is close to the optimal value $T_{\text{out}} \approx \mathcal{L} $. The cavity is kept on resonance with light using the Pound-Drever-Hall technique.  The number of atoms in the $F=4$ hyperfine ground state coupled to the light field (Fig.~\ref{fig:SpinOscillator}b) has been adjusted within the  $\sim 10^7-10^8$ range   by changing the cell temperature (typically $\sim 26^{\text{o}}$C) and optical pumping for the maximal QND interaction strength.
The frequency of the oscillator can be tuned with $B_0$.

\begin{figure}
  \includegraphics[scale=0.5]{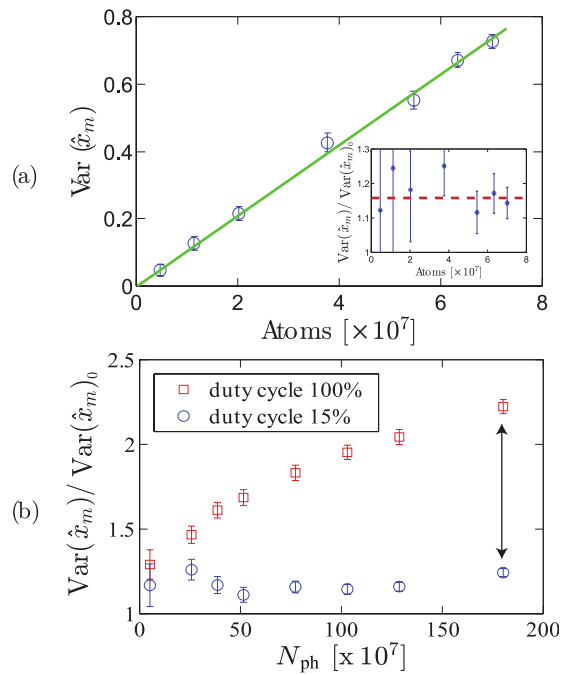}
  \caption{(a) Measured noise  in the oscillator state prepared by optical pumping. The noise is normalized to the probe light noise. Inset: Ratio of the measured oscillator variance to the expected measurement variance when the oscillator is in the ground state. The dashed line is the weighted average. (b) Demonstration of back-action noise suppression with stroboscopic quantum nondemolition measurement at twice the oscillator frequency. The oscillator noise has been normalized to the zero point fluctuations. Red squares - continuous measurement. Blue circles -  stroboscopic measurement with $15 \%$ duty cycle. The arrow indicates the back-action suppression.}
   \label{fig:SpinNoiseVsAt}
\end{figure}

An acousto-optic modulator is used to stroboscopically modulate the intensity of the probe beam at twice the Larmor frequency. The experiment was operated with $\approx 15 \%$ stroboscopic duty cycle, with probe wavelength blue detuned by 1.6 GHz with respect to the D2 transition \cite{Supplemental}. The $\hat{S}_{y,\tau}$ operator is measured by balanced polarimetry and lock-in detection. The data are weighted with an exponential mode function: $u(t) \propto e^{\pm \gamma t}$, where $\gamma$ is the decoherence rate in the presence of the probe. The exponential falling mode function is used to assess the measured noise, except for the squeezing investigation where the first pulse measurement is defined with a rising mode (see Fig.~\ref{fig:Experimental Setup}b). To collect statistics for the variance estimation, each measurement is repeated $\sim 2 \times 10^4$ times.

The atoms of the magnetic oscillator are initialized by optical pumping in a state close to the ground state.  The initial state variance is calibrated against a measurement of the noise in the thermal state of the atomic ensemble obtained with unpolarized spins.
The measured spin variance of unpolarized atoms $\propto N_{\text{at}}F(F+1)/3$ can serve as a robust reference for the spin noise in the coherent end state $\propto N_{\text{at}}F/2$ \cite{FPPRLNuclElectSq,RomalisStroboscopicPRL}, since the former is insensitive to classical fields and probe-induced noise.
In Fig.~\ref{fig:SpinNoiseVsAt}a the oscillator noise variance in the state prepared by optical pumping is plotted as a function of the atomic density. The observed linear scaling indicates a quantum-limited performance and a QND character of the measurement.
In the inset the ratio of the measured variance to the calibrated zero point imprecision is shown to be $\sim 1.16$. The increased measured variance in the initial oscillator state is due to the imperfect optical pumping, which leads to a finite oscillator temperature corresponding to thermal occupation $\bar{n} \approx \left( 8 \pm 1  \right) \times 10^{-2}$.
The occupation probability distribution among the Zeeman levels can be found from the magneto-optical resonance signal \cite{JulsgaardMORS} (see Fig.~3 in \cite{Supplemental}). Assuming a spin-temperature distribution, it is found that after optical pumping $ \sim 98 \%$ of the atoms are in the end state $|F, m_F \rangle = |4, -4 \rangle$, with $\sim 2 \%$ occupation probability for the $|F, m_F \rangle = |4, -3 \rangle$ state, and negligible probabilities for the other states. This is consistent with the 16\% increase of the measured variance compared to the ground state noise.

It is instructive to compare decoherence and thermalisation properties of the magnetic oscillator and mechanical oscillators. For the former oscillator initialized in the ground state (fully polarized spin), $\bar{n}(t)\propto f(F) e^{t/T_1}-1$ in the range of $t \leq T_1$ where $T_1$ is the population life time ($\sim 10$~msec) and $F/2 \leq f(F) \leq (F+1)(2F+1)/2$ in the present case. The thermalisation time $T_{\text{th}}= \frac{T_1 k_B T}{\hbar \Omega}$ is at least six orders of magnitude greater than $T_1$. For mechanical oscillators $\bar{n}(t)\propto \bar{n}_{\text{bath}} e^{t/T_{\text{th}}} \approx \bar{n}_{\text{bath}}$ at $t \approx T_{\text{th}}$ where the thermalisation time can reach at best a second in state-of-the-art experiments. This comparison illustrates the difficulty of quantum state preservation without cryogenic environment for mechanical oscillators \cite{PainterMechanicalCoolingNature} in comparison to magnetic spin oscillators.

\begin{figure}
  \includegraphics[scale=0.65]{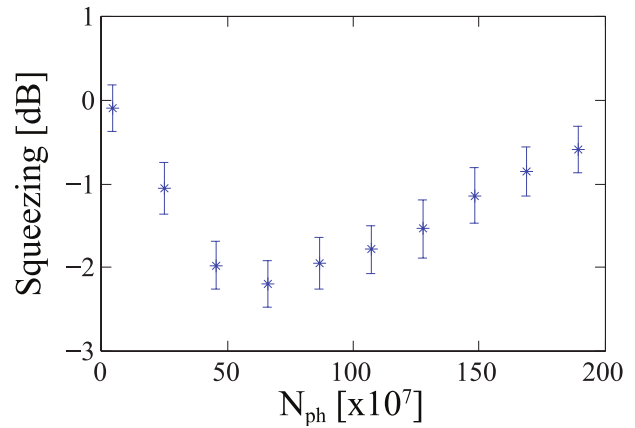}
  \caption{Conditional preparation of squeezed oscillator via stroboscopic QND measurement.}
   \label{fig:SpinSqueezing}
\end{figure}

The QND character of the measurement is demonstrated in Fig. \ref{fig:SpinNoiseVsAt}b, where the measured variance of the oscillator is plotted as a function of the number of the probe photons in the pulse. For a continuous probe ($100\%$ duty cycle), the imprecision in units of zero point fluctuations increases with the number of photons in the pulse.
In contrast, for a stroboscopic probe with a small duty cycle ($\sim 15\%$) the noise remains nearly independent of the pulse strength over the measured range.
The demonstrated reduction of probe back-action is more than 10 dB compared to a continuous probing of the oscillator.

To study the generation of squeezed oscillator states, two QND pulses are employed: the first provides information about the oscillator observable $\hat{x}$, and the second pulse evaluates the observable variance conditioned on the first measurement (see Fig.~\ref{fig:Experimental Setup}b).
In Fig.~\ref{fig:SpinSqueezing} the reduction in conditional variance compared to the zero point imprecision is plotted as a function of the number of photons in the first pulse, for a fixed photon number in the second pulse $N_{\text{ph,B}}\approx 27 \times 10^7$. The squeezing increases with the photon number in the pulse until the point where decoherence induced noise compensates for the reduction in uncertainty by measurement. The observed conditional variance is up to $2.2 \pm 0.2$ dB below the ground state noise.  Along with the dominant contribution of QND measurement to the degree of squeezing, a smaller contribution is due to the interaction nonlinear in the spin variables (see \cite{HapperOpticalPumping,Supplemental} for the effect of the second-rank tensor polarizability). The demonstrated noise reduction indicates the metrological advantage of the created state \cite{WinelandCriterion}.

In summary, we have demonstrated a squeezed state of an oscillator generated by a stroboscopic QND measurement. It has been made possible by implementing optical cavity-enhanced QND interaction and by utilizing a long lived spin oscillator with room temperature atoms contained in a spin protecting microcell. The techniques developed in the paper will be useful for quantum metrology and sensing, as well as for generation of entanglement between disparate oscillators \cite{HAPPRL09}.

This work was supported by the ERC grant INTERFACE, DARPA project QUASAR and EU grant SIQS. K. J. acknowledges support from the Carlsberg Foundation. G.V. gratefully acknowledges help and support from P. Karadaki.


%



\widetext
\clearpage
\begin{center}
\textbf{\large Supplementary information to ``Generation of a squeezed state of an oscillator by stroboscopic back-action-evading measurement"}
\end{center}
\setcounter{equation}{0}
\setcounter{figure}{0}
\setcounter{table}{0}
\setcounter{page}{1}
\makeatletter
\renewcommand{\theequation}{S\arabic{equation}}
\renewcommand{\thefigure}{S\arabic{figure}}
\renewcommand{\bibnumfmt}[1]{[S#1]}
\renewcommand{\citenumfont}[1]{S#1}

\section{Light-atomic spin interaction}
In this section we give a brief overview of the interaction between the light and spin atomic ensembles. We derive simple input-output equations describing the evolution of the light-oscillator system and relate the coupling constant to experimental parameters.

\begin{figure}
  \includegraphics[scale=0.6]{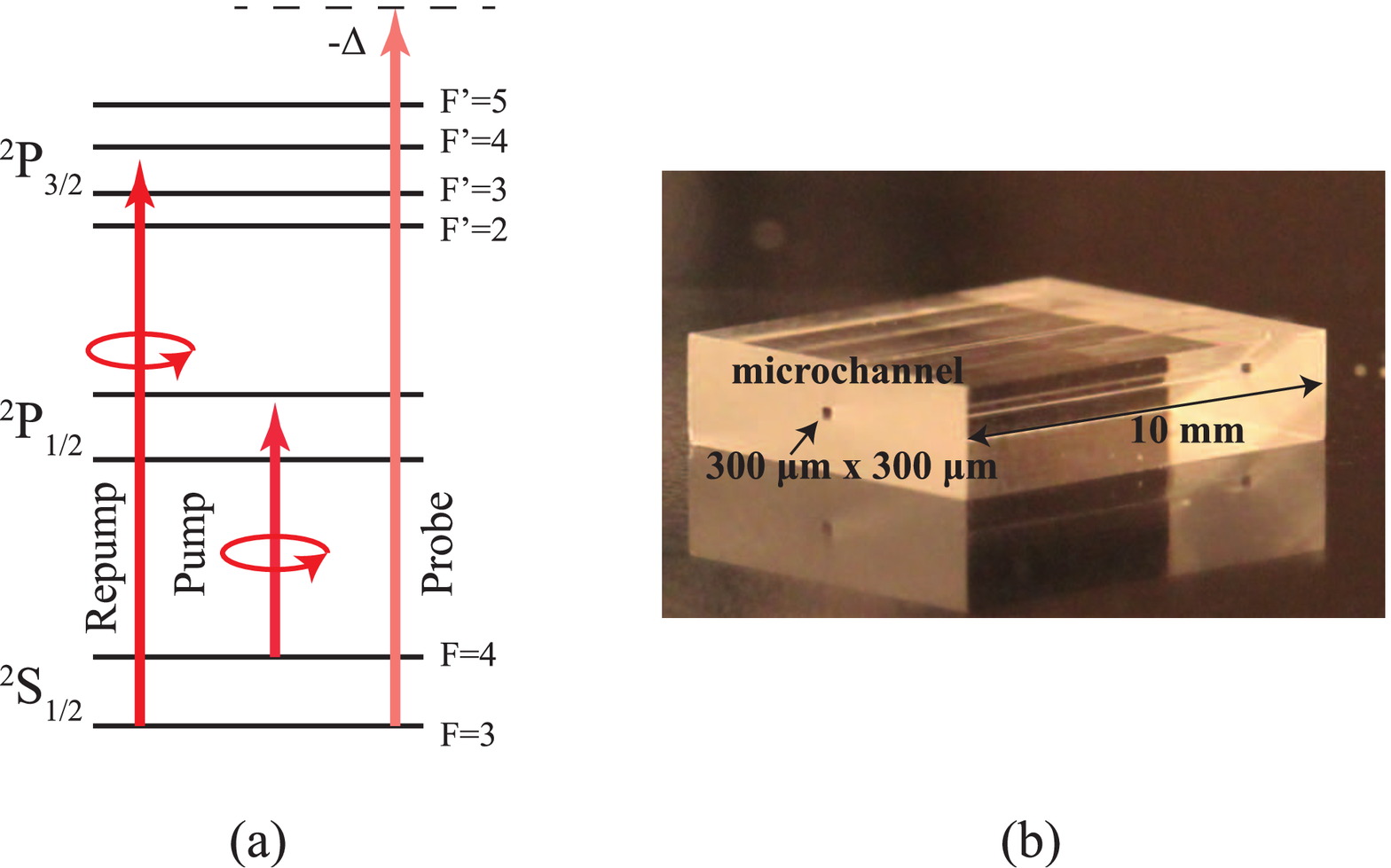}
  \caption{(a) Level structure of Caesium. The probe light is blue detuned by $-\Delta$ in the $^2S_{1/2} \rightarrow ^2P_{3/2}$ (D2 line), whereas the pump and repump light are on resonance with the D1 ($^2S_{1/2} \rightarrow ^2P_{1/2}$) and D2 line respectively. (b) The Cs atomic ensemble is contained in a glass microchannel 300 $\mu$m $\times$ 300 $\mu$m in cross section and 10 mm in length.}
 \label{fig:CsLevels}
\end{figure}

In the limit of low saturation parameter ($\frac{\chi^2}{ \Delta^2+\Gamma^2/4 } \ll1$, where $\chi$ is the Rabi frequency, $\Delta$ the probe detuning from the atomic resonance and $\Gamma$ is the FWHM homogeneous broadening of the electronic transition) the coherent interaction of light with atomic spin can be written in the form of a polarizability Hamiltonian \cite{SJulsgaardthesis, SKasperthesis} (throughout this letter we take $\hbar=c_L=1$, with $c_L$ being the speed of light):
\begin{equation}
H_0  =  - \frac{\Gamma}{8 A \Delta} \frac{\lambda^2}{2 \pi} \Bigg \{ a_0 +a_1 \hat{S}_z \hat{j}_z +a_2 \left[ \Phi \hat{j}_z^2-2\left(\hat{j}_x^2-\hat{j}_y^2 \right) \hat{S}_x -2 \left(\hat{j}_x \hat{j}_y+\hat{j}_y \hat{j}_x \right)\hat{S}_y \right]  \Bigg\}, \label{eq:Hamiltonian0}
\end{equation}
where $\lambda$ the wavelength of atomic transition, $A$ is the cross-sectional area of interaction, $\Phi$ is the photon flux, $\hat{j}_{x,y,z}$ are the components of the dimensionless atomic spin (sum of nuclear and electron spin), $a_{0,1,2}$ are the detuning dependent scalar, vector and second-rank tensor polarizabilities respectively. For the Cesium D2 transition ($6^2S_{1/2} \rightarrow 6^2P_{3/2}$) from the $F=4$ hyperfine manifold of the ground state, the polarizabilities are given by (see Fig.~\ref{fig:CsLevels}):
\begin{align}
a_0 & =\frac{1}{4} \left( \frac{1}{1-\Delta_{35}/\Delta}+\frac{7}{1-\Delta_{45}/\Delta}+8 \right),  \label{eq:ScalarPol}\\
a_1 & =\frac{1}{120} \left( -\frac{35}{1-\Delta_{35}/\Delta}-\frac{21}{1-\Delta_{45}/\Delta}+176 \right), \label{eq:VectroPol}\\
a_2 & =\frac{1}{240} \left( \frac{5}{1-\Delta_{35}/\Delta}-\frac{21}{1-\Delta_{45}/\Delta}+16 \right), \label{eq:TensorPol}
\end{align}
where $\Delta_{i5}$ is the frequency spacing between the $F'=i$ and the $F'=5$ excited states and $\Delta$ is the light detuning from the $F=4 \rightarrow  F'=5$ transition ($\Delta$ is negative for blue detuning). In Eq.~\ref{eq:Hamiltonian0} the operators $\hat{S}_{x,y,z}$ are the polarization Stokes components of light and can be described in terms of creation (annihilation) operators of the light field in the transverse polarizations. Assuming light propagation along the $z$ direction:
\begin{align}
\hat{S}_x &= \frac{1}{2} \left( \hat{a}_x^{\dag} \hat{a}_x - \hat{a}^{\dag}_y \hat{a}_y\right), \\
\hat{S}_y &= \frac{1}{2} \left( \hat{a}_x^{\dag} \hat{a}_y + \hat{a}^{\dag}_y \hat{a}_x\right), \\
\hat{S}_z &= \frac{1}{2 i } \left( \hat{a}_x^{\dag} \hat{a}_y - \hat{a}^{\dag}_y \hat{a}_x\right),
\end{align}
where the raising and lowering field operators $\hat{a}_{x,y}^{\dag}$, $\hat{a}_{x,y}$ are defined so that in the vacuum state $\left[ \hat{a}(t),\hat{a}^{\dag}\right]=\delta (t-t')$ and $\hat{a}^{\dag}(t) \hat{a}(t)$ is the flux of photons at time $t$. The operator $\hat{S}_z$ measures the light polarization in the circular basis, whereas $\hat{S}_{x,y}$ measure the polarization in complementary linear basis and are normalized so that for a coherent field linearly polarized in the $x$ direction: $\langle \hat{S}_y(t) \hat{S}_y(t') \rangle = \langle \hat{S}_z(t) \hat{S}_z (t') \rangle =(S_x/2) \delta (t-t') $ and $\langle \hat{S}_x \rangle = S_x = \Phi/2$.

In describing polarization spectroscopy, the scalar part of the Hamiltonian can be neglected, since it does not affect the evolution of atomic spin and the Stokes operators. For detunings much larger than the hyperfine splitting of the excited state $| \Delta | \gg (\Delta_{25},\Delta_{35},\Delta_{45}) $:  $a_2 \mapsto 0 $.
When the light interacts with an ensemble of $N_{\text{at}}$ atoms, all prepared in the same state and coupled with the same strength to the light mode, the coherent interaction of a light pulse of duration $\tau$ with the ensemble can be described  in the limit of large detuning  with the Hamiltonian:
\begin{align}
H & =   - \frac{ \Gamma}{8 A \Delta} \frac{\lambda^2}{2 \pi} a_1 \hat{S}_z \hat{J}_z= 2 \kappa \frac{\hat{S}_z}{\sqrt{N_{\text{ph}} }} \frac{\hat{J}_z}{\sqrt{F N_{\text{at}}}} \phantom{a}, \label{eq:HamiltonianKappa}  \\
\kappa & = - \frac{ \Gamma}{16 A \Delta} \frac{\lambda^2}{2 \pi} a_1 \sqrt{ N_{\text{ph}} F N_{\text{at}}} \phantom{a} , \label{eq:Kappa}
\end{align}
where $\hat{J}_z$ and $J_x$ refer to the ensemble collective spin components and $N_{\text{ph}}=\Phi \tau$ is the number of photons in the pulse.

Room-temperature atomic ensembles may introduce two complications concerning the interaction Hamiltonian: Doppler broadening and thermal motion in and out of the beam. For the experiment of this work, the detuning ($|\Delta| = 1600 $ MHz) was much larger than the Doppler width ($\Delta \nu_D \approx 200 $ MHz), so that the Doppler broadening has a small effect on the observed noise; this can also be seen from the fact that for a detuning much larger than the Doppler broadening the Voigt profile of the atomic line approximates a Lorentzian. Similarly, for the experimentally relevant time scales of the atomic state evolution the thermally moving atoms cross the beam many times and the effect of motion averages out \cite{SPokzikRMP}. In this case,  $A$ in Eq.~\ref{eq:HamiltonianKappa}-\ref{eq:Kappa} refers to the cross sectional area of the atomic cloud.

In the following, we will assume coherent linearly polarized light field in the $x$ direction probing a highly polarized atomic ensemble experiencing a DC magnetic field $B_0$ along the $x$ direction. For highly polarized atomic ensembles, the transverse collective spin components can be mapped to canonical position and momentum variables of an oscillator:
\begin{equation}
\hat{X} = \frac{\hat{J}_z}{\sqrt{|J_x|}}, \phantom{aaa} \hat{P} = \frac{\hat{J}_y}{\sqrt{|J_x|}}. \label{eq:XPDefine}
\end{equation}
The ground state of the oscillator corresponds to the coherent spin state of maximum spin in the $x$ direction; in this state $\langle \hat{X}^2 \rangle_0 = \langle \hat{P}^2 \rangle_0 =  1/2$.
Neglecting retardation effects in the light propagation and assuming without loss of generality $|J_x|=J_x$, the Hamiltonian evolution of $\hat{X}\propto \hat{J}_z$ and $\hat{P}\propto \hat{J}_y$ are described by the following equations \cite{SPolzikMultimodeEntanglementPRA}:
\begin{align}
\hat{X}(t) = & \cos ( \Omega t) \hat{X}(0) + \sin ( \Omega t) \hat{P}(0) + \beta \sqrt{J_x} \int_0^t \! \mathrm{d}t' \, \sin \left[ \Omega (t-t') \right] \hat{S}_z (t'), \label{eq:JzEvol} \\
\hat{P}(t) = &-\sin ( \Omega t) \hat{X}(0) + \cos ( \Omega t) \hat{P}(0) + \beta \sqrt{J_x} \int_0^t \! \mathrm{d}t' \, \cos \left[ \Omega (t-t') \right] \hat{S}_z (t'), \label{eq:JyEvol}
\end{align}
where $\Omega$ is the Larmor angular frequency $\Omega = \gamma_g B_0$, with $\gamma_g$ being the atomic gyromagnetic ratio and $\beta$ is a parameter defined by:
\begin{equation}
\beta =  - \frac{ \Gamma}{8 A \Delta} \frac{\lambda^2}{2 \pi} a_1 \phantom{a} . \label{eq:betaDef}
\end{equation}
Due to the Hamiltonian interaction with the atomic ensemble the $\hat{S}_y$ light operator evolves as:
\begin{equation}
\hat{S}_y^{\text{out}}(t)  = \hat{S}_y^{\text{in}}(t)+ \beta S_x \sqrt{J_x} \hat{X}(t), \label{eq:SyOut0}
\end{equation}
where the (in), (out) superscripts denote the input and output of the ensemble.

\section{Stroboscopic Squeezing}
In this section, we will estimate the squeezing that can be realized using stroboscopic probing of the oscillator. Initially, decoherence will be neglected (we will assume only Hamiltonian dynamics); the effect of decoherence will be considered at the end of the section. In the Hamiltonian dynamics, we will neglect the effect of second rank polarizability.

We consider the experimentally relevant case of linearly polarized probe beam in the $y$ direction, with stroboscopic intensity-modulation at twice the spin oscillator frequency; the Stokes element $\hat{S}_y^{\text{out}}$ is detected and the Fourier component $\cos (\Omega t)$ is measured by a lock-in amplifier.
In this way, the atomic-ensemble variable that is measured is $\hat{X}(t)$ integrated over one oscillator cycle and weighted with a cosine wave (from the lock-in amplifier demodulation) and a pulse-shaped function from the stroboscopic modulation of the probe:
\begin{equation}
\hat{x}(k T) = \frac{1}{T D} \int_{k T}^{(k+1)T} \! \mathrm{d}t \, \hat{X}(t) \phi(t) \cos (\Omega t), \label{eq:Defx}
\end{equation}
where $k$ is an integer number ($k \in \mathbb{N}$),
$T=2 \pi/\Omega$ is the oscillation period and the stroboscopic function $\phi(t)$ with duty cycle $D$ is given by (see Fig.~2~(b) in \cite{SMainPaper}):
\begin{equation}
   \phi(t) = \left\{
     \begin{array}{lrc}
       1 & :& -\frac{D T}{4}+k T \leq t \leq \frac{D T}{4}+k T \\
       0 & :& \frac{D T}{4}+k T < t < -\frac{D T}{4}+(k+1/2) T \\
       1 & :& -\frac{D T}{4}+(k+1/2) T \leq t \leq \frac{D T}{4} + (k+1/2)T
     \end{array}
   \right. , \label{eq:PhiDefinition}
\end{equation}
The variable $\phi(t)$ is defined so that the overlap with the lock-in cosine quadrature is maximized. The position variable $\hat{p} (k T)$ can be defined in the same way as in Eq.~\ref{eq:Defx} by replacing $\hat{X}$ with $\hat{P}$.

The squeezing $\xi^2_0$ realized after a measurement time $\tau =  N_{\text{m}} T $, with $ N_{\text{m}} \in \mathbb{N} $ being the number of complete oscillation cycles in the measurement, is given by:
\begin{equation}
\xi^2_0 = \frac{\text{Var} \Big[ \hat{x} ( N_{\text{m}} T ) \Big | \hat{S}_{y,\tau}^{\text{out}} \Big]}{\text{Var} \big[ \hat{x}\big]_0} ,  \label{eq:Xi0}
\end{equation}
where $\text{Var} \big[ \hat{x}\big]_0$ is the noise in the ground state of the oscillator (coherent spin state of maximum spin component along the $x$ direction) and $\hat{S}_{y,\tau}^{\text{out}}$ is the polarimetry measurement record.
For the experiment of this work (large number of atoms and photons), $\hat{x}(t)$ and $\hat{S}_{y,\tau}^{\text{out}}$ can be considered as Gaussian variables. Therefore, the conditional variance in Eq.~\eqref{eq:Xi0} can be written as \cite{SPokzikRMP}:
\begin{equation}
\text{Var} \Big[ \hat{x} ( N_{\text{m}} T ) \Big | \hat{S}_{y,\tau}^{\text{out}} \Big] = \text{Var} \Big[ \hat{x} ( N_{\text{m}} T )\Big]-\frac{\text{Cov}^2 \Big[ \hat{x} ( N_{\text{m}} T ) , \hat{S}_{y,\tau}^{\text{out}} \Big]}{\text{Var} \Big[ \hat{S}_{y,\tau}^{\text{out}} \Big]} \phantom{a}. \label{eq:CondVarGaussian}
\end{equation}

\subsection{Ground state imprecision}
In the oscillator ground state: $\hat{X}(t)=\hat{X}_0 \cos(\Omega t)+\hat{P}_0 \sin (\Omega t) $, with $\text{Var} \left( \hat{X}_0 \right)=\text{Var} \left( \hat{P}_0 \right) =1/2$, so that:
\begin{equation}
 \text{Var} \big[ \hat{x}\big]_0 = \frac{\left[ 1+\text{Sinc} (\pi D) \right]^2 }{8} \, . \label{eq:GroundStateNoise}
\end{equation}

\subsection{Measured Polarimetry noise}
For simplicity, we will consider  no weighting of the data with a mode function.
The measurement record: $\hat{S}_{y,\tau}^{\text{out}}= \int_0^{\tau} \! \mathrm{d}t \, \hat{S}^{\text{out}}_y(t) \cos (\Omega t) $ can be written in the form:
\begin{align}
S_{y,\tau} &= \sum_{k=0}^{N_{\text{m}}} \left[ \hat{Y}(k T)+ \beta \bar{S}_x T \sqrt{J_x} \hat{x}(k T) \right], \label{eq:MeasRecord} \\
\hat{Y}(k T) & = \int_{k T}^{(k+1)T} \! \mathrm{d}t \, \hat{S}_y^{\text{in}}(t) \phi(t) \cos (\Omega t) \,, \label{eq:YDef}
\end{align}
where the bar denotes average over the oscillator period. The first sum in Eq.~\ref{eq:MeasRecord} leads to the photon shot noise, while the second sum has the information about the oscillator state.

Since there are no correlations between the light at the input of the ensemble ($\hat{S}_y^{\text{in}}$) and the oscillator: $\langle \hat{x}(k_1 T ) \hat{Y}(k_2 T) \rangle=0, \phantom{a} \forall k_1,k_2$. The correlation related to the photon shot noise of light is:
\begin{equation}
\langle \hat{Y} (k_1 T) \hat{Y} (k_2 T) \rangle = \frac{\bar{\Phi} T}{8}\left[1+\text{sinc} \left( \pi D \right) \right] \delta_{k_1 k_2} = \langle \hat{Y}^2 \rangle_0 \delta_{k_1 k_2} \,. \label{eq:YYCor}
\end{equation}
with  $\delta$ being the Kronecker delta and $\bar{\Phi}$ the average photon flux over the oscillator period.

The variable $\hat{x}(k T)$ can be written as the sum of two uncorrelated contributions: $\hat{x}(k T) = \hat{x}_{\text{in}}(k T)+\hat{x}_{\text{BA}} (k T)$, with $\langle \hat{x}_{\text{in}}(k_1 T) \hat{x}_{\text{BA}} (k_2 T) \rangle=0, \phantom{a} \forall k_1,k_2 $.
The term $\hat{x}_{\text{in}}(k T)$  is related to the initial (before the interaction with the probe light) quantum state of the oscillator, which is assumed to be the ground state:
\begin{align}
\hat{x}_{\text{in}}(k T) & = \frac{1}{T D} \int_{k T}^{(k+1)T} \! \mathrm{d}t \, \phi(t) \cos (\Omega t) \left[ \hat{X}(0) \cos (\Omega t) +\hat{P}(0) \sin (\Omega t )  \right]  =\frac{ \hat{X}_0 \left[ 1+\text{sinc} (\pi D) \right] }{2} \, . \label{eq:NoiseTem1}
\end{align}
The term $\hat{x}_{\text{BA}} (k T)$  gives rise to the noise that describes the coupling of the quantum probe noise to the measured variable $\hat{x}$:
\begin{align}
\hat{x}_{\text{BA}} (k T) = \frac{1}{T D} \int_{k T}^{(k+1)T} \! \mathrm{d}t \, \phi(t) \cos (\Omega t) \hat{X}_{\text{BA}}(t),  \label{eq:BAN0}\\
\hat{X}_{\text{BA}} = \beta \sqrt{J_x} \int_0^t dt' \sin \left[ \Omega (t-t') \right] \hat{S}_z(t'). \label{eq:BAN1}
\end{align}

It can be shown that:
\begin{equation}
\langle \hat{x}_{\text{BA}}(k_1 T) \hat{x}_{\text{BA}}(k_2 T) \rangle = \frac{\left[ \mathcal{K}+2 \text{min}(k_1,k_2) \right] \beta^2 J_x  \bar{\Phi} T }{64} \big[ 1-\text{sinc}(\pi D) \big] \big[ 1+\text{sinc}(\pi D) \big]^2, \label{eq:XCor2}
\end{equation}
where  $\mathcal{K}$ is a numerical factor of order unity.
Here, we will consider the case where we are measuring over many cycles, so that $\mathcal{K}\ll N_{\text{m}}$. In this case, the measurement variance is:
\begin{align}
\text{Var} \Big[ \hat{S}_{y,\tau}^{\text{out}} \Big] &= N_{\text{m}}\langle \hat{Y}^2 \rangle_0+ \frac{\beta^2 J_x \bar{\Phi}^2  N_{\text{m}}^2 T^2}{4} \text{Var} \big[ \hat{x} \big]_0 + \frac{\beta^2 J_x \bar{\Phi}^2  T^2}{4} \displaystyle \sum_{k_1=0}^{N_{\text{m}}-1} \sum_{k_2=0}^{N_{\text{m}}-1} \langle \hat{x}_{\text{BA}}(k_1 T) \hat{x}_{\text{BA}}(k_2 T) \label{eq:VarSy1a} \rangle,  \\
& \approx \frac{\bar{\Phi} \tau}{8} \left[  1+\text{sinc}(\pi D) \right] \left[ 1+\tilde{\kappa}^2+\frac{\tilde{\kappa}^4}{3} \frac{1-\text{sinc}(\pi D)}{ 1+\text{sinc}(\pi D) } \right], \label{eq:VarSy1}
\end{align}
where the effective coupling constant is:
\begin{align}
\tilde{\kappa}^2 & =\frac{1}{4}\beta^2 J_x \bar{\Phi} \tau \left[  1+\text{sinc}(\pi D) \right]. \label{eq:K2SDef}
\end{align}

\subsection{Conditional variance of the oscillator observable}
The covariance between the quantum observable at the end of the measurement ($\hat{x}(N_{\text{m}} t)$) and the measurement record ($\hat{S}_{y,\tau}^{\text{out}}$) is:
\begin{align}
 \text{Cov}\Big[ \hat{x} ( N_{\text{m}} T ) , \hat{S}_{y,\tau} \Big] &= \frac{\beta \sqrt{J_x} \bar{\Phi} T }{2}  \Big[ N_\text{m}\text{Var} \big[ \hat{x}_0 \big] +\displaystyle \sum_{k=0}^{N_{\text{m}}-1}  \langle \hat{x}_{\text{BA}} (N_{\text{m}} T) \hat{x}_{\text{BA}} (k T) \rangle \Big], \label{eq:CovxSya} \\
  & \approx \frac{\beta \sqrt{J_x} \bar{\Phi} \tau}{16} \left[  1+\text{sinc}(\pi D) \right]^2 \left[ 1+ \frac{\tilde{\kappa}^2}{2} \frac{ 1-\text{sinc} (\pi D)}{1-\text{sinc} (\pi D)}  \right] \label{eq:CovxSyb}.
\end{align}
The unconditional variance of $\hat{x}(N_{\text{m}} t)$ is:
\begin{align}
\text{Var} \big[ \hat{x} (N_{\text{m}} T) \big] &= \text{Var} \big[ \hat{x}\big]_0 + \langle \hat{x}_{\text{BA}} (N_{\text{m}} T) \hat{x}_{\text{BA}} (kN_{\text{m}} T) \rangle , \label{eq:VarxNTa} \\
& \approx \text{Var} \big[ \hat{x}\big]_0\left[ 1+\tilde{\kappa}^2 \frac{ 1-\text{sinc} (\pi D)}{1-\text{sinc} (\pi D)} \right] . \label{eq:VarxNTa}
\end{align}
From Eqs.~\ref{eq:CondVarGaussian}, \ref{eq:K2SDef}, \ref{eq:CovxSyb} and \ref{eq:VarxNTa}, we get:
\begin{equation}
\text{Var} \Big[ \hat{x} ( N_{\text{m}} T ) \Big | \hat{S}_{y,\tau}^{\text{out}} \Big] = \text{Var} \big[ \hat{x}\big]_0 \Bigg[ 1+\tilde{\kappa}^2 \frac{ 1-\text{sinc} (\pi D)}{1-\text{sinc} (\pi D)}  -\frac{\tilde{\kappa}^2 \left[ 1+ \frac{\tilde{\kappa}^2}{2} \frac{ 1-\text{sinc} (\pi D)}{1-\text{sinc} (\pi D)}  \right]^2}{1+\tilde{\kappa}^2+\frac{\tilde{\kappa}^4}{3} \frac{1-\text{sinc}(\pi D)}{ 1+\text{sinc}(\pi D) }} \Bigg] \, . \label{eq:CondVarxNTSy}
\end{equation}

\subsection{Squeezing}
Combining Eqs.~\ref{eq:Xi0} and \ref{eq:CondVarxNTSy} we find that the squeezing is given by:
\begin{align}
\xi^2_0 \approx 1+\tilde{\kappa}^2 \frac{1-\text{sinc}(\pi D)}{ 1+\text{sinc}(\pi D) }-\frac{\tilde{\kappa}^2 \Big [ 1+\frac{\tilde{\kappa}^2}{2} \frac{1-\text{sinc}(\pi D)}{ 1+\text{sinc}(\pi D) } \Big]^2}{1+\tilde{\kappa}^2+\frac{\tilde{\kappa}^4}{3} \frac{1-\text{sinc}(\pi D)}{ 1+\text{sinc}(\pi D) }}. \label{eq:SqueezingFormula0}
\end{align}
In the limit of zero duty cycle ($D \rightarrow 0$): $\xi^2_0=1/(1+\tilde{\kappa}^2)$.

So far, it has been assumed that the oscillator does not experience any decoherence. The presence of decoherence results in reduction of the realized squeezing \cite{SPokzikRMP}:
\begin{equation}
\xi^2 \approx \xi_0^2+\eta_{\tau}, \label{eq:SqueezingFormula1}
\end{equation}
where $\eta_{\tau}$ scales with the decoherence events during the measurement time $\tau$ and depends on the decay mechanism. The decoherence can be linked to the light-probe, or it can be associated with coupling to a bath present even in the absence of the probe (e.g. coupling to the thermal phonon bath of a mechanical oscillator, or spin decay in the dark for a spin oscillator). In \cite{SFaradayNoiseDenis} exact relations for the probe induced decoherence and associated noise have been derived for the case of a spin oscillator. It is shown there that the probe induced noise can be written as:
\begin{equation}
\eta_{\tau} \big |_{\text{pr}} = \zeta \frac{\tilde{\kappa}^2}{d} \label{eq:ProbeDecNoise}
\end{equation}
where $\zeta$ is a numerical factor of order unity that depends on the probe polarization and the light-atom detuning, and ${d=n_{\text{at}} \sigma_0 l}$ is the optical depth on resonance, with $n_{\text{at}}$ being the atomic density,  $\sigma_0$  the absorption cross section on resonance and $l$ the length of the cell.

\section{Cavity enhancement of spin squeezing}

We consider a standing wave cavity with input (output) mirror power transmission coefficient $T_{\text{in}}$ ($T_{\text{out}}$), round-trip intracavity power loss $\mathcal{L}$ and atomic single-pass absorption $\alpha$. For high Finesse ($\mathscr{F} \gg 1$), the cavity power transmission coefficient on resonance is: $4 T_{\text{in}} T_{\text{out}}/(T_{\text{in}}+T_{\text{out}}+\mathcal{L}+2 \alpha)$, which in the limit of $2 \alpha \ll (T_{\text{in}}+T_{\text{out}}+\mathcal{L})$ can be written as:  $4 T_{\text{in}} T_{\text{out}}/(T_{\text{in}}+T_{\text{out}}+\mathcal{L}) \left[ 1- (2\mathscr{F}/\pi) \alpha \right]$,  where $\mathscr{F} \approx 2 \pi /(T_{\text{in}}+T_{\text{out}}+\mathcal{L})$ is the cavity Finesse. The single pass atomic absorption  $\alpha$ is enhanced by the cavity by the factor $2\mathscr{F}/\pi$. From Kramers-Kronig relations the atomic phase shift, and therefore the polarization rotation angle, is enhanced by the same factor.

Effectively, the cavity modifies the input-output relations Eq.\ref{eq:JzEvol}-\ref{eq:SyOut0} by enhancing the light-matter coherent interaction by $2\mathscr{F}/\pi$. If $\kappa_0$ is the coupling constant in the absence of cavity, then for the same number of detected photons in the pulse the cavity-enhanced coupling constant is:
\begin{equation}
\kappa_c = \frac{2\mathscr{F}}{\pi} \kappa_0  . \label{eq:KCavity}
\end{equation}
Equation~\ref{eq:SqueezingFormula0} for the conditional squeezing (in the absence of decoherence) should be modified accordingly.

\begin{figure}[htbp]
\begin{centering}
  \includegraphics[scale=1]{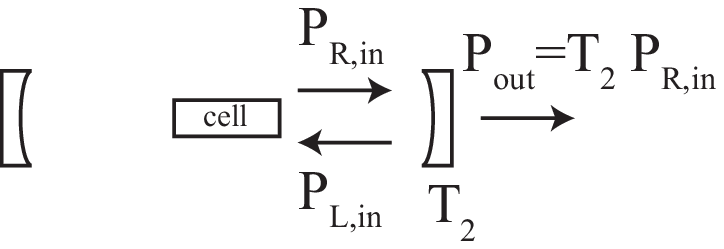}
   \caption{In a standing wave Fabry-Perot cavity there are leftwards (subscript L) and rightwards (subscript R) travelling fields. Assuming no losses in the cavity mirrors the following relations hold for the power of light field: $P_{\text{out}}=T_2 P_{\text{R,in}}$, $P_{\text{L,in}}=(1-T_2) P_{\text{R,in}}$, where the subscripts in (out) denote field in (out in the detection side) the cavity and $T_2$ is the power transmission coefficient of the output coupler. The total power experienced by the atomic ensemble contributing to decoherence due to spontaneous emission is: $P_{\text{in}}=P_{\text{R,in}}+P_{\text{L,in}}$.}
   \label{fig:CavityFields}
\end{centering}
\end{figure}

The realized squeezing is limited by the noise due to decoherence events. The probe-induced decoherence noise is proportional to the intracavity light-power experienced by the atomic ensemble $P_{\text{in}}$, which is related to the power at the output of the cavity ($P_{\text{out}}$) through: $P_{\text{out}} = P_{\text{in}} (2/T_2-1)$, where $T_2$ is the output mirror coupler power transmission (see Fig.~\ref{fig:CavityFields}). For a standing wave cavity, maximum squeezing is achieved for $T_1 \ll T_2$, so that the quantum light field generated inside the cavity from the interaction with the atoms has very small losses to the unobserved direction. Assuming only probe-induced decoherence and for $T_1 \ll T_2 \ll 1$, $\xi^2$ can be written in the form:
\begin{equation}
\xi^2 = \frac{1}{1+ \left( 2 \mathscr{F} / \pi \right)^2\kappa_0^2}+\eta_0 \left (\frac{2}{T_2}-1 \right) \approx \frac{1}{1+ \left( 2 \mathscr{F} / \pi \right)^2\kappa_0^2}+\frac{\mathscr{F} (T_2+\mathcal{L})}{\pi T_2} \eta_0, \label{eq:SqzCavity}
\end{equation}
where $\eta_0$ is the decoherence noise due to the light-probe in the absence of the cavity.
For small but finite intracavity losses, optimizing with respect to the number of photons in the light pulse and the output coupler transmission, the squeezing is:
\begin{equation}
\xi^2_{\text{opt}} \approx \left[ 2 - \sqrt{\frac{\zeta}{(2 \mathscr{F} / \pi) d_0}} \, \right] \sqrt{\frac{\zeta}{(2 \mathscr{F} / \pi) d_0 }} \approx \sqrt{\frac{\zeta}{(\mathscr{F} /2 \pi) d_0 }} \phantom{a}, \label{eq:SqzOptimal}
\end{equation}
which is realized for $T_2 = \mathcal{L}$. From the above equation, it can be seen that the presence of the cavity effectively enhances the optical depth by $\mathscr{F} /(2 \pi)$.

\section{Effect of second rank polarizability}
Equations~\ref{eq:JzEvol}-\ref{eq:SyOut0} describe the oscillator-light system evolution in the limit of zero second-rank tensor polarizability: $a_2 \rightarrow 0$. More complete equations, including the second-rank tensor polarizability effect, have been derived in \cite{SKasperthesis}. It can be shown that the \emph{Hamiltonian} evolution of the spin-oscillator can be approximated by:
\begin{align}
\hat{X}(t) &= \hat{X}(0) e^{-\gamma_{\text{sw}}t} \cos \left(\Omega t \right)+ \hat{P}(0)  e^{-\gamma_{\text{sw}}t} \sin \left(\Omega t \right) \nonumber \\
&+ \beta w \sqrt{J_x} \int_0^t \! \mathrm{d}t' \, e^{-\gamma_{\text{sw}}(t-t')} \cos \left[ \Omega (t-t')\right] S_y(t')  +  \beta \sqrt{J_x} \int_0^t \! \mathrm{d}t' \, e^{-\gamma_{\text{sw}}(t-t')} \sin \left[ \Omega (t-t')\right] S_z(t')\,, \label{eq:XEvolTensor} \\
\hat{P}(t) &= - \, \hat{X}(0)v   e^{-\gamma_{\text{sw}}t} \sin \left( \Omega t \right)+ \hat{P}_0  e^{-\gamma_{\text{sw}}t} \cos \left(\Omega t \right) \nonumber \\
&-  \beta w \sqrt{J_x} \int_0^t \! \mathrm{d}t' \, e^{-\gamma_{\text{sw}}(t-t')} \sin \left[ \Omega (t-t')\right] S_y(t')  +  \beta \sqrt{J_x} \int_0^t \! \mathrm{d}t' \, e^{-\gamma_{\text{sw}}(t-t')} \cos \left[ \Omega (t-t')\right] S_z(t') \,, \label{eq:PEvolTensor}
\end{align}
where for the Cs spin oscillator of the experiment: $w=14 a_2/a_1$, and the dissipation rate $\gamma_{\text{sw}}$, associated with the coherent interaction of the collective atomic ensemble spin with the light, is given by:
\begin{equation}
\gamma_{\text{sw}}= -\frac{S_x}{|S_x|} w \beta^2 N_{\text{at}} \Phi  \,. \label{eq:SwapRate}
\end{equation}
The parameter $\gamma_{\text{sw}}$ can either be negative or positive depending on the polarization direction of the linearly polarized probe light. In the experiment, a linearly polarized light in the $y$ direction was employed, resulting in $\gamma_{\text{sw}}$ being positive.

The effect of second rank tensor polarizability on the measured probe light Stokes component $\hat{S}_y^{\text{out}}$ can be neglected to first order in $w$ \cite{SKasperthesis}, and Eq.~\ref{eq:SyOut0} does not need to be modified: $\hat{S}_y^{\text{out}}(t)  = \hat{S}_y^{\text{in}}(t)+ \beta S_x \sqrt{J_x} \hat{X}(t) + \mathcal{O} \left( w^2 \right)$.

Through the interaction parameterized by $a_2$, noise in $\hat{S}_y^{\text{in}}$ couples into the measured observable $\hat{x}$ even in the limit of zero duty cycle. However, this back-action noise is suppressed by $w^2$ compared to the back-action noise arising from the vector polarizability Hamiltonian. The data presented in this work were acquired with a probe detuning $\Delta = -1.6$~GHz, leading to $w^2 < 10^{-2}$, so that the effect of tensor polarizability in the back-action noise can be ignored (its value is on the order of the back-action noise due to the finite duty cycle of the stroboscopic measurement).

It can be seen from the Eq.~\ref{eq:XEvolTensor} that the tensor polarizability dynamics can result in noise suppression for positive $\gamma_{\text{sw}}$ (the case of our experiment). This dissipative interaction can be used  to generate unconditionally squeezed states and steady state entanglement \cite{SPolzikDissipationPRL, SDenisDissipativeSqueezing}. In the experiment, second-rank tensor polarizability dynamics lead to small but finite reduction in the oscillator noise (see discussion about Fig.~\ref{fig:SupSqNoise}).

\section{Experimental setup}
The Cesium atomic ensemble is contained in a glass cell microchannel, shown in Fig.~\ref{fig:CsLevels}(b) filled with Cs vapor and with the walls covered with an alkene coating \cite{SBalabasPRLCoating}.
The cell is enclosed in a larger glass container.
The longitudinal spin lifetime is $T_1 \approx 17$~ms and the transverse spin lifetime is $T_2 \approx 10$ ms in the absence of light fields. The cell temperature (and the atomic density) can be regulated by using electrical heating.
The microcell is placed inside a standing wave optical cavity consisting of two concave mirrors. A mirror with $R_2=80\%$ reflectivity in intensity is used as the output coupler, while the input coupler has a much higher reflectivity: $R_1>99.7 \%$ in intensity. The cavity is locked on resonance by a Pound-Drever-Hall technique, applying a 10kHz modulation to the piezoelectric transducer. A digital PI controller, which incorporates feed-forward techniques, is employed to lock the cavity on resonance. The curvature (radius $r=110$ mm) and the separation of the mirrors are chosen so that the resulting cavity mode has $\approx 110$ $\mu$m waist diameter. This mode has small clipping losses in the cell, and allows for efficient coupling to the atomic ensemble. Due to the small clipping losses in the cell the cavity mode is Gaussian to a good approximation. The cavity Finesse is $\mathscr{F} \approx 17$, with a FWFM linewidth $\Delta \nu \approx 40 $ MHz.

Optical pumping of the ensemble is realized with circularly polarized pump and repump light fields resonant with the D1 and D2 atomic transition respectively (see Fig.~\ref{fig:CsLevels}(a) ). The repump light effectively transfers population from the $F=3$ hyperfine manifold to the probed $F=4$ manifold of the ground electronic state. The pump light field creates high degree of spin orientation in the $F=4$ manifold.

\section{Characterization of experimental parameters}

\subsection{Atomic Orientation}
\begin{figure}[htbp]
\begin{centering}
  \includegraphics[scale=0.8]{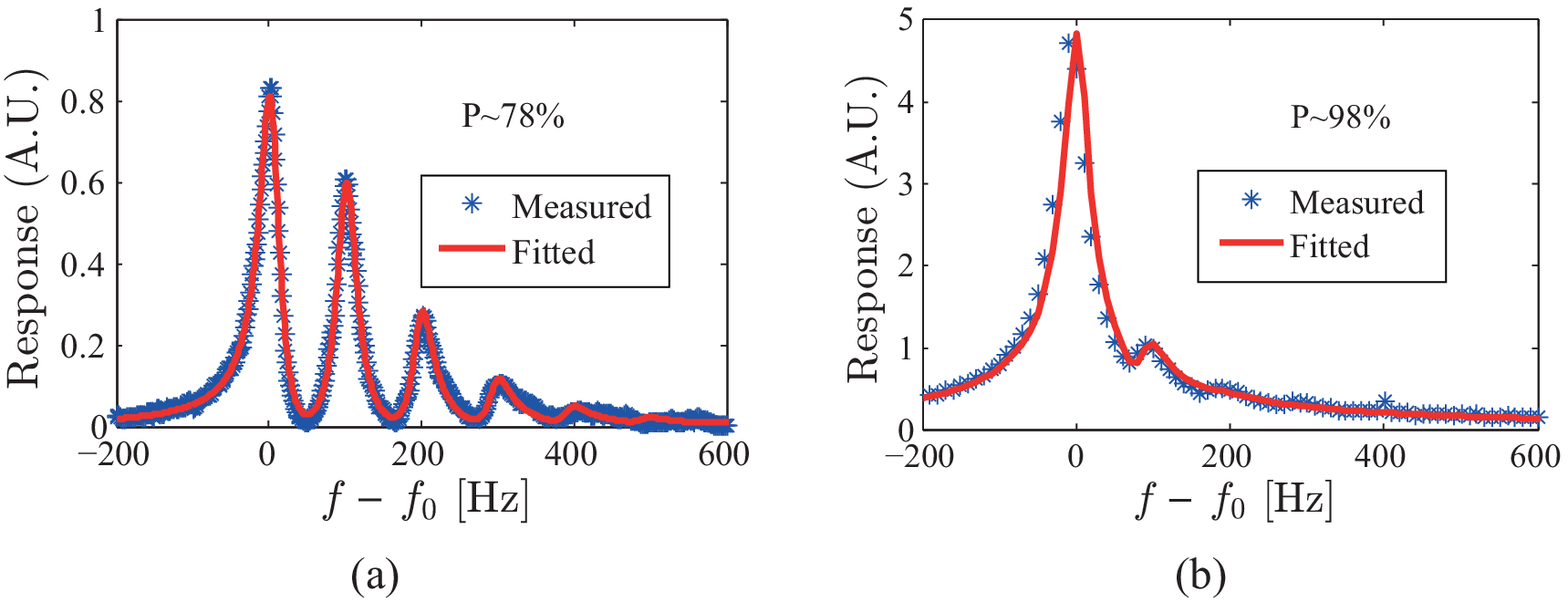}
   \caption{Fourier transform (amplitude) of the spin response to a short RF pulse for low (a) and high (b) spin orientation. The applied DC magnetic field is strong enough so that different Zeeman coherences have distinct frequencies due to the nonlinear Zeeman splitting. The bias frequency $f_0$ was close to 1~MHz. The spin orientation is estimated by fitting the experimental data to the spin temperature distribution.}
   \label{fig:MORS}
\end{centering}
\end{figure}

The atomic polarization can be estimated by mapping the RF resonance at large magnetic fields, where the non-linear Zeeman effect splits the magnetic resonances. A RF pulse, short enough so that the Fourier spectrum is approximately flat over the frequency range of the Zeeman resonances, is applied at the end of the pumping pulse and the spin evolution is monitored with a weak probe. In Fig.~\ref{fig:MORS} the RF resonance (Fourier transform of the spin response to the RF pulse) is plotted for two different degrees of spin orientation. Under the assumption of spin temperature distribution \cite{SHapperSpinTemperature}, which maximizes the entropy under the constraint of a given spin orientation, the spin ordination can be estimated by fitting the experimentally observed data to the model described in \cite{SPolzikJOptBMors}.
We note that for a weak probe the main spin-relaxation mechanism comes from the collisions with the walls of the cell and magnetic field gradients, so that a common relaxation rate for the ground state coherences can be assumed.

In order to estimate the fraction of atoms in the $F=3$ manifold due to imperfect optical pumping, the spin responses to short RF field resonant to the $F=3$ manifold Zeeman frequency in the cases of optical pumping phase with and without the repump light field are compared. We note that Cs has different gyromagnetic ratios in the two ground state manifolds resulting in different Larmor frequencies. In the absence of repump light during the pumping interval, the pump light depopulates almost completely the $F=4$ manifold creating orientation in the $F=3$ manifold. For this measurement, the probe wavelength is tuned to be mostly sensitive to transitions from the $F=3$ manifold. It is estimated that less than 5\% of the atoms remain in the $F=3$ manifold after the optical pumping in the experiment, so that for the prediction of the measured uncertainty in the oscillator ground state approximately all atoms in the ensemble should be considered.

\subsection{Number of atoms in the ensemble}
The number of atoms in the cell can be determined from the Faraday optical rotation of a linearly polarized probe traveling through the atomic ensemble that has been polarized along the light propagation direction. For a fully polarized ensemble (no atoms in the $F=3$ manifold and 100\% orientation in the $F=4$ manifold), the number of atoms $N_{\text{at}}$ is related to the Faraday angle $\theta_F$ by the equation:
\begin{equation}
N_{\text{at}} = \left| \frac{8 \pi A \theta_F \Delta}{a_1 \Gamma \lambda^2} \right| \,.
\end{equation}
The atomic density can be also characterized by optical absorption spectroscopy.
Typically, the agreement on atomic density between the Faraday angle measurement and the optical absorption spectroscopy is  $\sim 30\%$.

In Fig.3(a) of \cite{SMainPaper}, the number of atoms was calibrated against a response to a RF field for the polarized atoms and independently by a spin noise measurement for the unpolarized atoms. The atomic response to a RF field much shorter than the spin coherence time is $\propto N_{\text{at}}$ for a fixed spin orientation, so that by monitoring the relative response to a given RF field the relative number of atoms can be found. Similarly, the spin noise of atoms in the thermal state (where no probe back-action occurs) scales proportionally to the number of atoms.

\subsection{Photon shot noise level}
The photon shot noise level can be estimated from a polarimetry noise measurement when the oscillator frequency is tuned away from the lock-in detection bandwidth. The noise associated with the spin-oscillator has a Lorentzian power spectrum centered at Larmor frequency with width set by the decoherence rate; therefore if the oscillator frequency is set many line-widths away from the lock-in bandwidth, the oscillator noise is filtered out and the measurement noise only depends on the probe light noise of white spectrum. The oscillator frequency can be tuned by changing the bias magnetic field. The probe noise is verified to scale linearly with the power as expected for the photon shot noise power.

\subsection{Electronic noise and detection losses}
The photodetector electronic noise is at the level of $\sim 10 \%$ of the photon shot noise for the power used in the squeezing experiment. The detection efficiency of the light field at the output of the cavity is $\sim 0.85$, limited mainly by the detector quantum efficiency.

\subsection{Ground state noise}
The ground state noise is calibrated against a measurement in the thermal state of the atomic ensemble. Including decoherence, but neglecting change in the spin variance (e.g. due to loss of atoms from depumping into the $F=3$ manifold), the two-time spin correlation can be written in the form: $\langle J_z(t) J_z(t') \rangle = \langle J_z^2 \rangle e^{- \gamma |t-t'|} \cos \left[ \Omega (t -t') \right]$, where $\gamma$ is the decay rate and $\langle J_z^2 \rangle $ is the variance in the spin state. It can be shown that the $\cos \left( \Omega t \right)$ component of the $\hat{S}_y^{\text{out}}$ measurement over time $\tau$, weighted by the exponential mode function $u(t) \propto e^{-\gamma_{\text{m}} t}$, is given (in the absence of back-action) by:
\begin{equation}
\frac{\text{Var} \left[ \int_0^{\tau} \! \mathrm{d}t \, \hat{S}_y^{\text{out}} (t) \cos \left( \Omega t \right) u(t) \right]}{ \text{PSN}_{\tau}} = 2 \beta^2 S_x \frac{\gamma -e^{2  \gamma_{\text{m}} \tau} \gamma+\left[1+e^{2 \gamma_{\text{m}} \tau}-2 e^{(\gamma_{\text{m}}-\gamma) \tau}\right] \gamma_{\text{m}} }{\left(-1+e^{2 \gamma_{\text{m}} \tau}\right) (\gamma_{\text{m}}-\gamma ) (\gamma+\gamma_{\text{m}})}\langle \hat{J}_z^2\rangle \, , \label{eq:VarJzDec0}
\end{equation}
which for $\gamma_{\text{m}} = \pm \gamma $ reduces to:
\begin{equation}
 \frac{\text{Var} \left[ \int_0^{\tau} \! \mathrm{d}t \, \hat{S}_y^{\text{out}} (t) \cos \left( \Omega t \right) u(t) \right]}{ \text{PSN}_{\tau}}=  \beta^2 S_x \tau \frac{1+ \gamma \tau - \gamma \tau  \coth ( \gamma \tau )}{\gamma \tau} \langle \hat{J}_z^2 \rangle \, , \label{eq:VarJzDec1}
\end{equation}
where $\text{PSN}_{\tau}$ is the photon shot noise (variance) contribution to the measurement noise, estimated by:
\begin{equation}
 \text{PSN}_{\tau}=\int_0^{\tau} \! \mathrm{d}t \, \hat{\tilde{S}}_y^{\text{out}} (t) \cos \left( \Omega t \right) u(t) \,, \label{eq:PSNestimation}
\end{equation}
with $\hat{\tilde{S}}_y^{\text{out}}$ being the polarimetry output when the oscillator frequency is detuned away from the lock-in bandwidth. In the thermal state, where no back-action occurs, the spin variance is: $\langle \hat {J}_z^2 \rangle|_{\text{th}} = \frac{2F+1}{4 F} N_{\text{at}} F (F+1)/3$, where $F=4$ is the total spin in the probed manifold, $N_{\text{at}}$ is the total number of atoms in the ensemble and the factor $(2F+1)/(4 F)$ is the fraction of atoms in the $F=4$ manifold. We note that in the experiment described here, the contribution of atoms in the $F=3$ manifold to the thermal spin noise measurement is estimated to be $<3\%$. The ground state of the oscillator corresponds to the coherent spin state of maximum $J_x=N_{\text{at}} F$, leading to $\langle \hat{J}_z^2 \rangle =N_{\text{at}} F/2$. Assuming the same decoherence rate for coherent and thermal states, we find that the contribution of a magnetic oscillator in the ground state with $J_x=N_{\text{at}} F/2$ to the measured noise is given by (in units of $\text{PSN}_{\tau}$) :
\begin{equation}
\text{Var} \left( \hat{x}_m \right)_0 =  \frac{6 F}{(F+1)(2F+1)} \left[  \frac{\left. \text{Var} \left[ \int_0^{\tau} \! \mathrm{d}t \, \hat{S}_y^{\text{out}} (t) \cos \left( \Omega t \right) u(t) \right] \right|_{\text{thermal}} }{ \text{PSN}_{\tau}}-1 \right] \,. \label{eq:GroundNoiseCal}
\end{equation}
There is a small discrepancy in the decoherence rate and the loss rate of atoms due to the probe scattering for the coherent and thermal rates; for the timescales of the experiment, the effect of this difference is estimated to be $<10\%$.

\section{Squeezing}
In order to study squeezing, a scheme with two successive pulse measurements is employed (see Fig.~2(b) in \cite{SMainPaper}). The first measurement is described by:
\begin{equation}
\hat{q}_A = \sqrt{\frac{4 \gamma}{e^{2 \gamma \tau_A}-1}} \int_0^{\tau_A} \! \mathrm{d}t \, \hat{S}_y^{\text{out}}(t) \cos \left( \Omega t \right) e^{\gamma t}\,, \label{eq:SqueezA}
\end{equation}
while the second measurement is:
\begin{equation}
\hat{q}_B = \sqrt{\frac{4 \gamma}{1-e^{-2 \gamma \tau_B}}} \int_{\tau_A}^{\tau_A+\tau_B} \! \mathrm{d}t \, \hat{S}_y^{\text{out}}(t) \cos \left( \Omega t \right) e^{-\gamma t}\,. \label{eq:SqueezB}
\end{equation}
An exponentially rising (falling) mode is used for the first (second) pulse to moderate the effect of  decoherence which reduces the correlations between the pulses.

The measurement noise associated with the oscillator state is extracted by subtracting the photon shot noise for the same pulse duration ($\text{PSN}_{A,B}$) from the measured noise; normalized to the photon shot noise variance the measured oscillator noise is:
\begin{align}
\text{Var} (\hat{x}_{m,A}) & =  \left[ \frac{\text{Var} (\hat{q}_{A})}{\text{PSN}_{A}}-1 \right]\,, \label{eq:XmVarA} \\
\text{Var} (\hat{x}_{m,B}) & = \left[ \frac{\text{Var} (\hat{q}_{B})}{\text{PSN}_{B}}-1 \right]\,, \label{eq:XmVarB} \\
\text{Var} (\hat{x}_{m,B|A}) & = \left[ \frac{\text{Var} (\hat{q}_{B}|\hat{q}_{A})}{\text{PSN}_{B}}-1 \right]= \left[ \frac{\text{Var} (\hat{q}_{B})}{\text{PSN}_{B}} - \frac{\text{Cov}^2 \left[ \hat{q}_{B},\hat{q}_{A}\right]}{\text{PSN}_B \text{Var}\left( \hat{q}_{A} \right)}-1 \right]\,. \label{eq:XmVarCond}
\end{align}
The conditional variance $\text{Var} (\hat{x}_{m,B|A})$ effectively evaluates the oscillator position noise in the state filtered out by the first measurement. The realized squeezing is determined by the measurement strength in the first pulse; in Eq.~\ref{eq:SqueezingFormula0} and Eq.~\ref{eq:SqueezingFormula1} the relevant coupling constant $\kappa$ is associated with the first measurement.

\begin{figure}
\begin{centering}
  \includegraphics[scale=0.8]{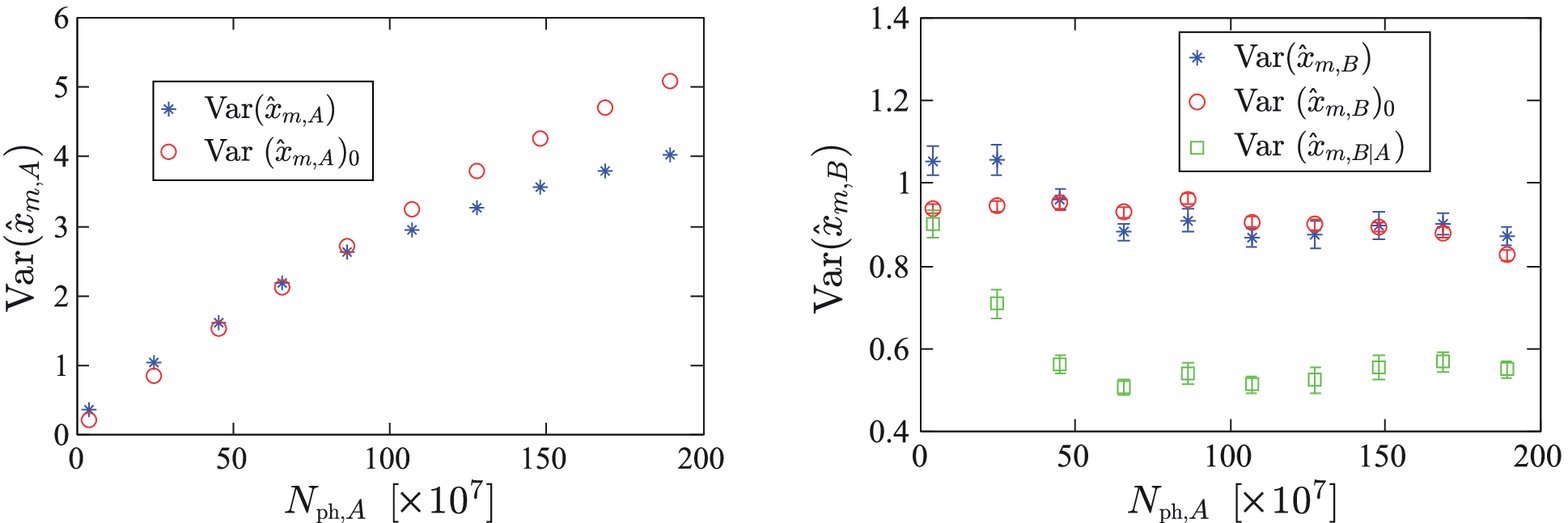}
   \caption{(a) Measured noise ($\text{Var} (\hat{x}_{m,A})$) and expected noise in the oscillator ground state  ($\text{Var} (\hat{x}_{m,A})_0$)  as a function of the number of photons $N_{{\text{ph}},A}$. (b) Comparison of expected ground state noise ($\text{Var} (\hat{x}_{m,B})_0$) with unconditional and conditional to the first pulse measured noises ($\text{Var} (\hat{x}_{m,B})$ and $\text{Var} (\hat{x}_{m,A})_0$ respectively) as a function of the number of photons in the first pulse. See text for a discussion. }
   \label{fig:SupSqNoise}
\end{centering}
\end{figure}

In Fig.~\ref{fig:SupSqNoise} the oscillator noise in the first and second pulse, conditionally and unconditionally to the first measurement, are plotted as a function of the number of photons in the first pulse $N_{{\text{ph}},A}$. In the experiment, the probe power averaged over an oscillator period is maintained constant and the duration of the first pulse ($\tau_A$) is varied. The duration of the second pulse is fixed to 0.5 ms, which corresponds to $\sim 27 \times 10^7$ photons. The expected noise for the ground state of the oscillator (denoted by the subscript 0) is also plotted for comparison. Due to the correlation between the first and the second pulse, the conditional variance of the second pulse is significantly reduced below the position imprecision in the oscillator ground state, leading to squeezing. The data in Fig.~\ref{fig:SupSqNoise} indicate an additional squeezing mechanism unrelated to the information gain by measurement, since the unconditional noise for both the first and second pulse drops below the ground state uncertainty. This kind of squeezing is due to the second-rank tensor polarizability dynamics.

\begin{figure}
\begin{centering}
  \includegraphics[scale=0.8]{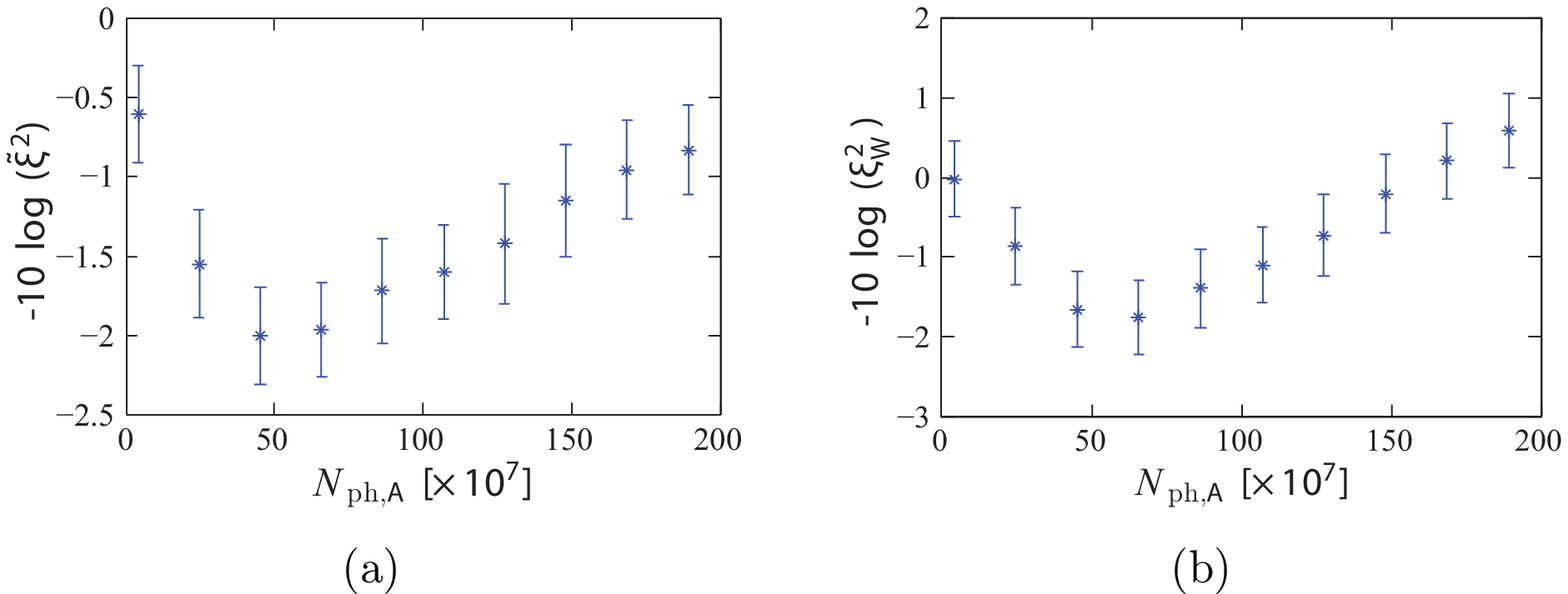}
   \caption{(a) Uncertainty reduction of the oscillator position  by using the information gained by a QND measurement.  (b) Demonstrated squeezing according to the Wineland criterion, which quantifies the spectroscopic sensitivity enhancement with respect to the sensitivity in the ground state of the oscillator.}
   \label{fig:SupSqzPlots}
\end{centering}
\end{figure}

In Fig.~\ref{fig:SupSqzPlots}(a) the conditional noise is evaluated with respect to the measured unconditional imprecision. A comparison of the data in Fig.~\ref{fig:SupSqzPlots}(a) with those in Fig.~4 of \cite{SMainPaper} indicates that the realized degree of squeezing is dominated by the information gain through the QND measurement.

In evaluating the squeezing degree, the measured variance is normalized to the macroscopic spin component $J_x  $ at the end of the first pulse: $\tilde{\xi}^2=  \text{Var} (\hat{x}_{m,B|A})/  \left(  \text{Var}  (\hat{x}_{m,B}) f_d \right) $, where $f_d=J_x (\tau_A)/(4 N_{\text{at}})$ accounts for the reduction of the mean spin during the first measurement. The same correction factor $f_d$ is used to estimate the squeezing in Fig.~4 of \cite{SMainPaper}. The reduction of the mean spin $J_x$ is characterized by an independent measurement of the response to an external coherent perturbation. The probe pulse of duration $\tau_A$, applied after the optical pumping phase, is followed by a resonant RF excitation pulse, much shorter than the decoherence time, and the response, which is proportional to the mean spin $J_x$, is recorded.

In Fig.~\ref{fig:SupSqzPlots}(b) the demonstrated squeezing  according to the Wineland criterion \cite{SWinelandCriterion} is plotted as a function of $N_{{\text{ph}},A}$. The Wineland squeezing evaluates the effect of noise suppression in spectroscopic sensitivity: $\xi_{\text{W}}^2= \text{Var} (\hat{x}_{m,B|A})/( f_d^2 \text{Var}  (\hat{x}_{m})_0)$. The data show that the created squeezed state offers the possibility of enhanced force measurement below the ground state uncertainty.


%

\end{document}